\documentclass[11pt,superscriptaddress,nofootinbib]{article}
\usepackage{cite}
\usepackage{amsmath,amsfonts,amssymb}
\usepackage[small,bf,hang]{caption}
\usepackage[bookmarks=true,hyperfigures=true]{hyperref}
\usepackage{slashed}
\usepackage{mathabx}
\usepackage{latexsym,epsfig}
\usepackage{color}
\usepackage{mathtools}
\usepackage{nicefrac}

\def\hybrid{
        \topmargin -20pt
        \oddsidemargin 0pt
        \headheight 0pt \headsep 0pt
        \textwidth 6.25in 
        \textheight 9.5in 
        \marginparwidth .875in
        \parskip 5pt plus 1pt \jot = 1.5ex}

\hybrid

 \csname
@addtoreset\endcsname{equation}{section}

\newcommand{\vev}[1]{\langle #1 \rangle}
\newcommand{\cPhi}{\check{\Phi}}
\newcommand{\eqn}[1]{(\ref{#1})}

\def\moth{\mathsurround=0pt}
\newdimen\zo \zo=0pt

\def\tick{\leaders\hrule height 0.5ex depth 0pt \hskip 0.5pt}
\def\upboxfill{$\moth \setbox\zo\hbox{\tick}%
  \hskip 3pt\hbox to 0pt{$\tick$\hss}\hrulefill \hbox to 7.5pt{$\tick$\hss}$}

\def\dtick{\leaders\hrule height .34pt depth 0.5ex \hskip 0.5pt}
\def\downboxfill{$\moth \setbox\zo\hbox{\dtick}%
  \hskip 2pt\hbox to 0pt{$\dtick$\hss}\hrulefill \hbox to 2pt{$\dtick$\hss}$}

\newcommand{\sfrac}[2]{{\textstyle\frac{#1}{#2}}}

\def\nl{\nonumber\\}
\def\bec{\begin{center}}
\def\ec{\end{center}}

\def\c{\gamma} 

\def\d{\delta}

\def\ve{\varepsilon}

\def\l{\lambda}

\def\r{\rho}
\def\s{\sigma}

\def\pa{{\partial }}
\def\Ra{{\Rightarrow}}

\def\bfR{{\bf{R}}}

\def\cD{{\cal D}}
\def\cC{{\cal C}}
\def\cP{{\cal P}}

\def\cO{{\cal O}}

\def\cN{{\cal N}}
\def\cR{{\cal R}}
\def\cS{{\cal S}}

\def\cO{{\cal O}}

\def\nn{\nonumber}

\def\Tr{{\rm Tr}\,}

 \def\det{{\rm det\,}}
\def\be{\begin{equation}}
\def\ee{\end{equation}}
\def\bea{\begin{eqnarray}}
\def\eea{\end{eqnarray}}
\def\ba{\begin{array}}
\def\ea{\end{array}}

\begin{document}


\begingroup\raggedleft\footnotesize\ttfamily
HU-EP-20/06
\vspace{15mm}
\endgroup

\begin{center}
{\LARGE\bfseries $\cN =4$ super Yang-Mills correlators without anti-commuting variables
\par}%

\vspace{15mm}

\begingroup\scshape\large 
Hermann Nicolai${}^{a}$ and Jan~Plefka${}^{b}$ 
\endgroup
\vspace{5mm}

\textit{${}^{a}$
Max-Planck-Institut f\"ur Gravitationsphysik
(Albert-Einstein-Institut)\\
M\"uhlenberg 1, D-14476 Potsdam, Germany
 } \\[0.25cm]
					
\textit{${}^{b}$Institut f\"ur Physik und IRIS Adlershof, Humboldt-Universit\"at zu Berlin,\\
  Zum Gro{\ss}en Windkanal 6, D-12489 Berlin, Germany} \\[0.25cm]

\bigskip
  
\texttt{\small\{nicolai@aei.mpg.de, jan.plefka@hu-berlin.de\}} 

\vspace{8mm}


\textbf{Abstract}\vspace{5mm}\par
\begin{minipage}{14.7cm}
Quantum correlators of pure supersymmetric Yang-Mills theories in $D=3,4,6$ and 10
dimensions can be reformulated via the non-linear and non-local transformation
(`Nicolai map') that maps the full functional measure of the interacting theory to
that of a free bosonic theory. As a special application we show that for the maximally 
extended $\cN=4$ theory in four dimensions, and up to order $\cO(g^2)$, all known results 
for scalar correlators can be recovered in this way without any use of anti-commuting 
variables, in terms of a purely bosonic and ghost free functional measure for the gauge fields. 
This includes in particular the dilatation operator yielding the anomalous dimensions
of composite operators. The formalism is thus competitive with more standard 
perturbative techniques. 
 \end{minipage}\par

\end{center}
\setcounter{page}{0}
\thispagestyle{empty}
\newpage


\section{Introduction}
Pure supersymmetric Yang-Mills theories exist in $D=3,4,6$ and 10 dimensions \cite{SYM1}.
As is well known, the corresponding extended super-Yang-Mills theories in lower dimensions 
can be obtained from these by dimensional reduction.  Among the supersymmetric Yang-Mills theories, 
the maximally extended $\cN=4$ theory in four dimensions stands out for several
reasons, especially in connection with the AdS/CFT correspondence, as a result of which 
there now exists an enormous variety and wealth of results (indeed, too many to list here!).
In this paper we want to take a new and different look at 
this theory, exploiting the existence of a non-local and non-linear transformation $T_g$
(`Nicolai map') that maps the full functional measure of the interacting Yang-Mills theory 
to the one of a theory of dim$\,G$ free (Maxwell) vector fields, where $G$ is the gauge group
in question (usually $G=SU(N)$). The existence of this map for the $\cN=1,D=4$ theory was 
established long ago \cite{Nic1,Nic2}, and a detailed prescription for its iterative construction
was presented in \cite{FL,DL1,DL2,L1} and \cite{Nic2}. It was, however, only very recently 
that these constructions were extended to other dimensions, and in particular to the maximally 
extended $D=10$ and $\cN=4,D=4$ theories \cite{ANPP}. The existence of the map $T_g$
opens very different perspectives on the quantization of supersymmetric Yang-Mills 
theories, in terms of a ghost and fermion free formalism and with a purely bosonic 
functional measure. This concerns especially the computation of quantum correlators.
Previous work in this direction remains somewhat scattered: in \cite{DL1,DL2} several perturbative
results for the $\cN=1,D=4$ theory (for instance, wave function renormalization factors
and the $\beta$-function to order $g^2$) were recovered in a perturbative approach.
Non-perturbative aspects were studied in \cite{Fub1} where it was shown in particular that 
there exists a {\em local} expression for $T_g$ in the light-cone gauge. This result was subsequently 
used to recalculate 2-gluon and 3-gluon Green's functions up to one loop \cite{FLMR}.
The $\cN=1,D=4$ Yang-Mills theory can also be investigated in terms of anti-selfdual variables, 
yielding (amongst other results) a non-perturbative derivation of  the $\beta$-function \cite{Boc1}.
However, as far as we are aware, \cite{Fub2} is the only attempt towards 
understanding {\em extended}, and more specifically, half-maximal 
({\em i.e.} $D=6$ or $\cN=2,D=4$) super-Yang-Mills theories in this framework, 
with an intriguing proposal for a closed form expression of $T_g$. Yet, to the best of
our knowledge, no results in this direction have been available so far for the maximally 
extended  $\cN=4$ theory, which from many points of view is by far the most interesting.
This is the main issue we want to (begin to) address in this paper.

Accordingly, we wish to investigate certain quantum correlators, and more specifically scalar
correlation functions of the $\cN=4$ theory in terms of the map $T_g$, and to show 
that several known results can be easily recovered with this formalism and in terms of
the map $T_g$, at least to the extent that it has been worked out. It should, however, be 
understood that these results -- being confined to the perturbative domain -- constitute only a 
very first step. Ultimately, we would hope that this formalism can provide essentially 
new insights on the $\cN=4$ theory. Amongst other 
things, these include prospects for a non-perturbative regularization of the $\cN=4$ theory, 
especially in conjunction with its conjectured integrability properties \cite{Beisert}.

The non-linear and non-local transformation (which more generally exists
for all rigidly supersymmetric theories with Lagrangians quadratic in the fermions)
\be
T_g[A]_\mu^a(x) \equiv A^{'a}_\mu(x,g;A)
\ee
is characterized by the following properties:

\begin{enumerate}
\item Substitution of $A'(A)$ into the free Maxwell action (or rather: sum of Maxwell actions)
yields the interacting theory, {\em viz.}
\be
\cS_0[A'(A)] = \cS_g[A] \equiv \frac14 \int d^D x\,  F_{\mu\nu}^a F_{\mu\nu}^a
\ee 
where
\be\label{Fmn}
F_{\mu\nu}^a \,\equiv \,  \pa_\mu A_\nu^a - \pa_\nu A_\mu^a + gf^{abc} A_\mu^b A_\nu^c
\ee
is the Yang-Mills field strength [with fully antisymmetric structure constants $f^{abc}$
for the chosen gauge group, usually SU($N$)], and $\cS_0$ is the free Maxwell action 
\be
\cS_0[A'] \equiv \frac14 \int d^D x\,  (\pa_\mu A_\nu^{\prime a} - \pa_\nu A_\mu^{\prime a})^{2}\, ,
\ee
i.e.~$\cS_g$ for $g=0$. 
\item $T_g$ preserves the gauge condition
\be 
T_g[G^a(A)] = G^a(A)
\ee
\item The Jacobian of the transformation equals the product of the Matthews-Salam-Seiler 
(MSS) determinant (or Pfaffian) \cite{MSS} obtained by integrating out the gauginos, 
and the Faddeev-Popov (FP) determinant \cite{FP} (obtained by integrating out 
the ghost fields $C^a,\bar{C}^a$), 
\be\label{Det}
\det \left(\frac{\delta A^{'a}_\mu(x,g;A)}{\delta A^b_\nu(y)}\right) = \Delta_{MSS} [A]\;\Delta_{FP}[A]
\ee
at least in the sense of formal power series.
\end{enumerate}

The existence of the map $T_g$ allows for a ghost free and fermion free quantization 
of supersymmetric  theories, and can thus provide a 
completely different perspective also on super-Yang-Mills theories. The main advance of 
the present work consists in applying these techniques to the computation of simple 
correlators for the maximal $\cN=4$ theory, and in showing that the calculational effort 
with this formalism is comparable to the usual one, thus providing a proof 
of principle for its workability and demonstrating its competitiveness with
more standard perturbative techniques. Of course, to push these computations further
one must determine the map $T_g$ to higher orders. Ultimately, the main goal
would be to go beyond the perturbative framework, by exploiting as yet unknown
properties of the map $T_g$, presumably related to the maximally extended
superconformal symmetry of the $\cN=4$ theory. Certainly it would be fascinating to make 
a connection between the map $T_g$ and the integrable properties of the $\cN=4$ theory 
(see e.g.~\cite{Beisert} for a review) -- after all the image of the map $T_{g}$
is a free field theory which is certainly integrable. A distinctive feature of the 
map $T_g$ is that it works for finite $N$ in the $SU(N)$ gauge theory, in contradistinction 
to integrability, which is tied to the planar ($N\to \infty$) limit. Indeed, while it appears unlikely that 
there exists a closed form expression for $T_g$ (as is the case for some special 
theories, like supersymmetric quantum mechanics and the $\cN=2,D=2$ Wess-Zumino 
model, see \cite{CG,Nic3,Fub1,L2}) there could be an underlying integrable structure.
Likewise, it would be interesting to find a link with the conformal bootstrap program 
(see {\em e.g.} \cite{Rychkov} for a review), where again the $\cN=4$ theory appears to play 
a distinguished role \cite{Osborn} (see also \cite{Rastelli} and references therein for more
recent work) and to elucidate the role of the conformal and dual-conformal symmetries in this context.

\section{Preliminaries}

Let us briefly summarize our conventions.  We use the Euclidean metric; this
is not essential, as analogous results can be derived
with Lorentzian signature (as in \cite{FL,DL1,DL2}).
The scalar propagator is (with the Laplacian $\Box \equiv \pa^\mu \pa_\mu$)
\be
C(x) = \int \frac{d^D k}{(2\pi)^D} \frac{e^{ikx}}{k^2} \quad \Ra \quad
- \Box C(x) = \d(x) 
\ee
where $\d(x) \equiv \d^{(D)}(x)$ is the $D$-dimensional $\d$-function. 
For the relevant dimensions we have
\bea\label{C}
C(x) &=&  \frac1{4\pi}\cdot \frac1{|x|}     \quad\,\qquad \mbox{for $D=3$} \nl
C(x) &=&  \frac1{4\pi^2}\cdot \frac1{x^2}     \quad\,\qquad \mbox{for $D=4$} \nl
C(x) &=&  \frac1{4\pi^3}\cdot \frac1{(x^2)^2}     \qquad \mbox{for $D=6$} \nl
C(x) &=&  \frac3{2\pi^5}\cdot \frac1{(x^2)^4}     \qquad \mbox{for $D=10$}
\eea
For all dimensions the free fermionic propagator is
\be
\c^\mu\pa_\mu S_0(x)   = \d (x)   \quad \Ra \quad S_0 (x) = - \c^\mu \pa_\mu C(x)
\ee
The number $r_D$ of 
spinor components depends on $D$, and we here restrict attention to those values for 
which supersymmetric Yang-Mills theories exist, {\em viz.}
\be\label{Dr}
D\,=\,  3,4,6,10  \qquad \Longleftrightarrow \qquad r_D\,=\,  2,4,8,16
\ee
For $D=4$ this corresponds to a Majorana spinor, for $D=6$ to a Weyl spinor, while
for $D=10$ we get an extra factor of $\frac12$ because of the 
Majorana-Weyl condition (otherwise we would have $r=32$). Here and in other 
formulas below we usually suppress spinor indices. To derive the extended theories
in four dimensions we will consider {\em dimensional reduction} of the corresponding 
theories to $D=4$ such that all integrals will be performed {\em in four dimensions}
(or rather, $D=4 - 2\ve$ for the regularized theory).

Covariant derivatives are only needed for the adjoint representation:
\be \label{Dmu}
D_\mu V^a \,\equiv \, \pa_\mu V^a + g f^{abc} A_\mu^b V^c  \;\; \Rightarrow \quad
[D_\mu , D_\nu] \, V^a = g f^{abc} F_{\mu\nu}^b V^c
\ee
Although results also hold for other gauges, we will here stick with the Landau 
gauge fixing function
\be
G^a[A_\mu] = \pa^\mu A_\mu^a 
\ee

For the map $T_g$ there is a systematic construction via its inverse $T_g^{-1}$ in terms of its infinitesimal generator \cite{FL,DL1,DL2,Nic2,L1}. The latter is realized by the so-called 
$\cR$-operator, such that
\be\label{Tginv}
(T_g^{-1}A)_\mu^a(x) \,=\, A_\mu^a(x) \,+ \,
\sum_{n=1}^\infty \frac1{n!}  \, g^n \, \Big(\cR^n  \big[ A \big]_\mu^a(x)\Big)_{g=0}
\ee
As we will see it is also the {\em inverse} map that is needed for the computation
of quantum correlation functions. For the Landau gauge the $\cR$-operator is 
compactly represented by the (functional) differential operator
\be\label{cR}
\cR = \frac{d}{dg}  - \, \frac1{2r_D} \int dx\,du\,dv\; 
\Pi_{\mu\nu}(x-u){\rm Tr}\, \big( \c_\nu \c^{\r\s} S^{ba}(v-u)\big) f^{bcd} A_\r^c(v)A_\s^d(v)
        \frac{\d}{\d A_\mu^a(x)}
\ee
 with the transversal projector
\be\label{Pi1}
\Pi_{\mu\nu}(x-y) \,\equiv \, \left( \d_{\mu\nu} - \frac{\pa_\mu \pa_\nu}{\Box}\right)\d(x-y)
\, \cong \, \d_{\mu\nu} \d(x-y) + \pa_\mu C(x-y) \pa_\nu 
\ee
where "$\cong$" means equality in the sense of distributions. 
Note that in the above we write  $du=d^{D}u$ for short and that space-time derivatives 
on propagators are to be understood as
$
\partial_{\mu}C(x-y):= \frac{\partial}{\partial x^{\mu}} C(x-y)
$, {\it i.e.}~as acting always on the first argument.
$S^{ab}(x,y;A)$ is the full
fermionic propagator in the gauge field dependent background with $A_\mu^a(x)$, 
and  thus defined by
\be
\c^\mu \big[ \d^{ac} \partial_\mu - gf^{acd} A_\mu^d(x)  \big] \, S^{cb}(x,y;A) = \d^{ab}\d(x-y)
\ee
The $\cR$ operator acts distributively, 
\be
\cR\big[A^a_\mu(x) A_\nu^b(y) \cdots \big] \, \,= \cR\big[A_\mu^a(x)\big] A_\nu^b(y)\cdots \, + \, 
    A_\mu^a(x) \cR\big[A_\nu^b(y)\big]\cdots \, + \, \cdots
\ee
Specializing the action of $\cR$ to the gauge field $A_\mu^a$, we get 
\be\label{bR1}
\cR[A]_\mu^a(x) \,\equiv\, - \, \frac1{2r_D} \int du dv\, 
\Pi_{\mu\nu}(x-u){\rm Tr}\, \big( \c_\nu \c^{\r\s} S^{ba}(v-u)\big)
      f^{bcd} A_\r^c(v)A_\s^d(v)
\ee
From (\ref{bR1}) it follows immediately that the $\cR$ operation preserves the Landau gauge
\be
\pa^\mu \, \cR\big[ A_\mu^a(x)\big] = 0 
\ee
This will guarantee that the equality
\be
\pa^\mu (T_g(A)_\mu^a) (x) = \pa^\mu A_\mu^a(x)
\ee
holds for all values of the Yang-Mills coupling constant $g$.  Once we have the result
for $T_g^{-1}$ the map $T_g$ itself can be obtained by perturbatively inverting the
power series (\ref{Tginv}) (in principle, there is also a direct construction of $T_g$ \cite{L2}).


To order $\cO(g^2)$ a double application of the $\cR$-operator leads to \cite{ANPP}
\begin{align}\label{Tginv}
(T_g^{-1} A)^a_\mu(x) \,&=\,  A_\mu^a(x) \, - \, g f^{abc} \int du\, \pa_\l C(x-u) A_\mu^b(u) A_\l^c(u)  \nl[1mm]
& \; + \,\frac12 g^2 f^{abc} f^{bde} \int dv dw\, 
    \Big[  -\pa_\r C(x-v) A_\s^c(v) \pa_\s C(v-w) A_\r^d(w) A_\mu^e(w) \nl
&  \qquad\qquad+ \pa_\r C(x-v) A_\s^c(v) \pa_\r C(v-w) A_\s^d(w) A_\mu^e(w) \nl[1mm]
& \qquad\qquad   -\pa_\r C(x-v) A_\s^c(v) \pa_\mu C(v-w) A_\s^d(w) A_\r^e(w) \nl[1mm]
& \qquad\qquad + 2 \,\pa_\r C(x-v) A_\mu^c(v) \pa_\s C(v-w) A_\s^d(w) A_\r^e(w) \nl[1mm]
& \qquad\qquad  -2 \, \pa_\r C(x-v) A_\r^c(v) \pa_\s C(v-w) A_\s^d(w) A_\mu^e(w) \Big]
  \; + \;\cO(g^3)
\end{align}
The map $T_g$ itself is obtained by inverting up to second order 
\bea\label{Tg}
(T_gA)^a_\mu(x) \,&=&\,  A_\mu^a(x) \, + \, g f^{abc} \int du\, \pa_\l C(x-u) A_\mu^b(u) A_\l^c(u)
 \\[2mm]
 &&  \hspace{-5mm} + \, \frac32 \, g^2f^{abc}f^{bde} \int du dv \, \pa_\r C(x-u) A_\l^c(u)
        \pa_{[\mu} C(u-v) A_\l^d (v) A_{\r]}^e(v)  \, + \, \cO(g^3) \nonumber
\eea
thus reproducing the old result from \cite{Nic1}. These formulas are valid 
in all dimensions where pure supersymmetric Yang-Mills theories exist.
While our main interest is in the maximally extended $\cN=4$ theory in 
four dimensions, we will keep $D$ general in the following section, and consider 
the dimensional reduction to $D=4$ in later sections.

\section{Correlation functions}
For all admissible dimensions, and for any $n$-point correlator of {\em bosonic} operators 
$\cO_j(x_j)$ our basic relation is
\be\label{correlator}
\big\langle \!\!\big\langle \cO_1(x_1) \cdots \cO_n(x_n) \big\rangle\!\!\big\rangle \,= \,
\big\langle T^{-1}_g[\cO_1](x_1) \cdots T^{-1}_g[\cO_n](x_n) \big\rangle_0
\ee
where $\cO_j(x_j)$ are either elementary or composite bosonic fields.
Here $\big\langle \!\!\big\langle \cdots \big\rangle\!\!\big\rangle$ denotes the full
expectation value  of the interacting supersymmetric Yang-Mills theory (with fermions, ghosts 
and all interactions), while $\big\langle\cdots\big\rangle_0$ denotes the free 
field expectation value of the purely bosonic non-interacting gauge theory where one integrates only 
over the bosonic fields (with the notation from \cite{DL1,DL2}). More precisely, we have 
\bea\label{corr0}
\big\langle \!\!\big\langle \cO_1(x_1) \cdots \cO_n(x_n) \big\rangle\!\!\big\rangle 
\,&\equiv&\, \int \cD A \cD \chi \cD C \cD \bar{C}  \, \prod_{x,a}  \delta\big( \partial^\mu A^a_\mu(x)\big) \,
e^{- S[A,\chi,C,\bar{C}]}\cO_1(x_1) \cdots \cO_n(x_n) 
\nl[1mm]
&=& \int \cD_g[A] \, \prod_{x,a}  \delta\big( \partial^\mu A^a_\mu(x)\big) \,  \cO_1(x_1) \cdots \cO_n(x_n) 
\eea
where $S[A,\chi,C,\bar{C}]$ is the full supersymmetric action (with gauginos $\chi^a$ 
and ghost fields $\{C^a,\bar{C}^a\}$), while $\cD_g[A]$ denotes the (non-local)
bosonic  functional measure of the interacting  theory obtained  {\em after integrating out 
the gauginos and the ghosts}. Likewise
\be\label{corr1}
\big\langle T_g^{-1}[O_1](x_1) \cdots T_g^{-1} [\cO_n](x_n) \big\rangle_0
\,\equiv \,
\int \cD_0[A] \, \prod_{x,a}  \delta\big( \partial^\mu A^a_\mu(x)\big)\,  
T_g^{-1}[O_1](x_1) \cdots T_g^{-1} [\cO_n](x_n) 
\ee
with the free measure $\cD_0[A]$ (where the fermionic and ghost determinants become trivial).
Importantly, the gauge fixing function is not affected by the transformation since 
$\cR(\partial^\mu A_\mu^a) = 0$ hence $\partial^\mu A^{' a}_\mu = \partial^\mu A^a_\mu $
to any given order. Due to the presence of the gauge fixing $\delta$-functional in (\ref{corr0}) 
the vector propagator is
\be\label{AA1}
\langle A^a_\mu(x) A^b_\nu(y) \rangle_0  \,= \,C^\perp_{\mu\nu}(x-y) \, \equiv \,
 \d^{ab} \left(  \d_{\mu\nu} - \frac{\pa_\mu \pa_\nu}{\Box}\right) C(x-y) 
\ee
For both the interacting and the free theory one can make use of the 't Hooft trick of 
shifting the argument of the $\delta$-functional by $c^a$ and integrating with a Gaussian 
weight over the dummy variable $c^a$ to remove the $\delta$-functional, and implement the 
gauge condition via the Gaussian factor 
$\propto \,\exp\left(- \frac{1}{2\xi}\int (\partial\cdot A)^2\right)$
in the functional integral, thereby introducing the gauge parameter $\xi$. While the Landau gauge
corresponds to $\xi=0$ we shall work here in the Feynman-gauge ($\xi=1$) for which 
the propagator takes the more convenient form
\be\label{AA2}
\big\langle A_\mu^a(x)\, A_\nu^b(y) \big\rangle_0 =
\delta^{ab}\, \delta_{\mu\nu}\, C(x-y) \, .
\ee
Equivalently, we can ignore the longitudinal contributions  as they will
drop out in all gauge invariant expressions. Below we will therefore use the
propagator in the form (\ref{AA2}).

Let us also note that with either choice, the free measure is already properly normalized 
for supersymmetric theories because
\bea
\int \cD A \, e^{\frac12 \int A\Box A} \,&\sim& \, [\det(-\Box)]^{-D/2} \; ,\nl[1mm]
\int \cD C \cD \bar{C} \, e^{\int \bar C \Box C} \,&\sim&\, \det (-\Box)\; , \nl[1mm]
\int \cD\chi \, e^{\frac12 \int \bar\chi \partial\!\!\!/ \chi} \,&\sim& \, [(\det(-\Box)]^{r_D/4}
\eea
if $r_D = 2(D-2)$,  which implies that bosonic and fermionic degrees of freedom match on shell.

In summary, by means of (\ref{correlator}) we are able to express any bosonic correlator of the fully
supersymmetric theory as a {\em purely bosonic} correlator with a purely bosonic functional
measure. In fact, the same statement also applies to fermionic correlators if we replace
the fermionic two-point functions by the full propagator $S(x,y;A)$ in a gauge field background
(and a $2n$-point correlator by the corresponding Wick product). Alternatively, one may
invoke supersymmetry to reduce fermionic correlators to bosonic ones via
superconformal Ward identities \cite{Osborn}.
Analogous relations also hold for composite operators, as we will illustrate below.

As a simple example we compute the 2-point function to second order
\be\label{2point}
\big\langle \!\!\big\langle A_\mu^a(x) A_\nu^b(y)\big\rangle\!\!\big\rangle \,=\,
\big\langle T_g^{-1}[A_\mu^a](x)T_g^{-1}[A_\nu^b](y) \big\rangle_0
\ee
At $\cO(g^2)$ there are two contributions from (\ref{Tginv}), namely the free Wick-contractions of the product of 
two $\cO(g)$ terms
\begin{align}
g^2 f^{acd} f^{bmn} &\int du dv \, \partial_\l C(x-u) \partial_\r C(y-v)  
\langle A_\mu^c(u) A_\l^d(u) A_\nu^m(v) A_\r^n(v) \rangle _0 \nn\\
=  g^2 N \d^{ab} & \int du dv \, \partial_\l C(x-u) \partial_\r C(y-v)  
\Big( C^\perp_{\mu\nu}(u-v) C^\perp_{\l\r} (u-v) - C^\perp_{\mu\r}(u-v) C^\perp_{\l\nu} (u-v) \Big) 
\end{align}
as well as the contractions emerging from the product of the leading order term with the order $\cO(g^2)$ terms
\begin{align}\label{Og2}
\frac12  g^2 N \d^{ab} \int du dv & \Big[ -\partial_\r C(x-u)  C^\perp_{\s\r}(u-v) \partial_\s C(u-v) C^\perp_{\mu\nu} (v-y) 
\nn\\
&  \; \; + \; \partial_\r C(x-u)  C^\perp_{\s\mu}(u-v) \partial_\s C(u-v) C^\perp_{\r\nu} (v-y) \nn\\
&  \; \; + \; \partial_\r C(x-u)  C^\perp_{\s\s}(u-v) \partial_\r C(u-v) C^\perp_{\mu\nu} (v-y) \nn\\
&  \; \; -  \; \partial_\r C(x-u)  C^\perp_{\s\mu}(u-v) \partial_\r C(u-v) C^\perp_{\s\nu} (v-y) \nn\\
&  \; \; - \; \partial_\r C(x-u)  C^\perp_{\s\s}(u-v) \partial_\mu C(u-v) C^\perp_{\r\nu} (v-y) \nn\\
&  \; \; + \; \partial_\r C(x-u)  C^\perp_{\s\r}(u-v) \partial_\mu C(u-v) C^\perp_{\s\nu} (v-y) \nn\\
&  \; \; + \; 2\, \partial_\r C(x-u)  C^\perp_{\mu\s}(u-v) \partial_\s C(u-v) C^\perp_{\r\nu} (v-y) \nn\\
&  \; \; - \; 2\, \partial_\r C(x-u)  C^\perp_{\mu\r}(u-v) \partial_\s C(u-v) C^\perp_{\s\nu} (v-y) \nn\\
&  \; \; - \; 2 \, \partial_\r C(x-u)  C^\perp_{\r\s}(u-v) \partial_\s C(u-v) C^\perp_{\mu\nu} (v-y) \nn\\
&  \; \; + \; 2\, \partial_\r C(x-u)  C^\perp_{\r\mu}(u-v) \partial_\s C(u-v) C^\perp_{\s\nu} (v-y) \Big]
     \;\; + \;\; (x\leftrightarrow y)
\end{align}
For the reasons explained above we can neglect longitudinal contributions,
and thus replace the transversal propagator $C^\perp_{\mu\nu}(x)$ by the  simpler
expression $\delta_{\mu\nu} C(x)$ in (\ref{AA2}). Then a straightforward calculation gives 
\begin{align}
\label{2pointA} 
\big\langle \!\!\big\langle A_\mu^a(x) A_\nu^b(y)\big\rangle\!\!\big\rangle \,=
&\,= \,
\delta^{ab}\, \delta_{\mu\nu}\, \Bigl [ C(x-y)  +g^{2}N\, \frac{6-D}{2}\, \int du \, C(x-u)\, C(y-u)^{2}\Bigr ] \nn\\
&-\delta^{ab}\, g^{2}N\, \frac{6-D}{2}\,\frac{\partial}{\partial x^{\mu}}\, \frac{\partial}{\partial y^{\nu}}\, \int dudv\,  C(x-u) \, C(y-v)\,
C(u-v)^{2}\, ,
\end{align}
where all that was used were suitable partial integrations. Recall that $du$ is a 
shorthand notation for the $D$-dimensional measure $d^{D}u$. 
Inspecting the integrands near four dimensions reveals a logarithmic divergence 
in both integrals when the argument of the squared Green's function $C(z)$ vanishes.
Therefore, the two-point function exhibits a divergence, illustrating the (known)
fact that `finiteness' of the theory does not mean that every correlator is finite.
Curiously, the $D=6$ theory, and thus also the $\cN=2$ theory in $D=4$, 
has a vanishing next-to-leading order contribution.

Similarly one may obtain the three-point function at the leading perturbative order 
upon expanding $T^{-1}_{g}$ to $\cO(g)$ in each term
\begin{align}
\label{3pointA}
\big\langle \!\!\big\langle A_{\mu_{1}}^{a_{1}}(x_{1}) A_{\mu_{2}}^{a_{2}}(x_{2})
A_{\mu_{3}}^{a_{3}}(x_{3})\big\rangle\!\!\big\rangle &=
\big\langle T_g^{-1}[A_{\mu_{1}}^{a_{1}}](x_{1})T_g^{-1}[A_{\mu_{2}}^{a_{2}}](x_{2})
T_g^{-1}[A_{\mu_{3}}^{a_{3}}](x_{3})\big\rangle_0 = \\ &
f^{a_{1}a_{2}a_{3}}\, g\, \Bigl [\delta_{\mu_{1}\mu_{2}}\, \left ( \frac{\partial}{\partial x^{\mu_{3}}_{2}}
 - \frac{\partial}{\partial x^{\mu_{3}}_{1}} \, \right ) 
 + \delta_{\mu_{2}\mu_{3}}\, \left ( \frac{\partial}{\partial x^{\mu_{1}}_{3}}
 - \frac{\partial}{\partial x^{\mu_{1}}_{2}} \, \right ) \nn \\ &
 + \delta_{\mu_{3}\mu_{1}}\, \left ( \frac{\partial}{\partial x^{\mu_{2}}_{1}}
 - \frac{\partial}{\partial x^{\mu_{2}}_{3}} \, \right )
\, \int du\,  C(x_{1}-u)\, C(x_{2}-u) \, C(x_{3}-u)
\Bigr ]\, , \nn
\end{align}
reproducing the standard three-gluon vertex of Yang-Mills theory. In order to 
compute the one-loop correction to this result we would need to 
know the map $T^{-1}_g$ to cubic order $\cO(g^{3})$.

\section{Scalar correlation functions at one loop in the $\mathcal{N}=4$ theory}

Next we turn to the extended theories in {\em four} dimensions which can be obtained
by dimensional reduction. To this aim we split the indices as $\mu \rightarrow \{ \mu,i \}$ 
where $\mu,\nu,...= 1,...,4$ and $i,j,...$ label the remaining internal dimensions. 
Likewise we decompose the coordinates as $x^\mu \rightarrow \{ x^\mu,  y^i\}$
and the fields and indices in (\ref{Tginv}) in an analogous fashion:
\be\label{dimred}
A_\mu^a (x,y) \;\; \longrightarrow \;\; \big\{ A_\mu^a(x)  \,,\, \phi_i^a (x) \big\}
\ee
The dependence on the internal coordinates $y^i$ is dropped for the dimensionally reduced theory.
We then proceed to compute the scalar two and four-point functions up to the next-to-leading 
perturbative order in the gauge coupling constant $g$. In the remainder we shall
focus on the $\cN=4$ super Yang-Mills theory, for which the internal indices run over six
dimensions: $i,j,\dots =1,\dots,6$. In the reduced and regulated  theory all loop integrals are 
performed in $D= 4- 2\ve$ dimensions while the number of scalars is $S=6+2\ve$. This prescription
maintains the balance
of fermionic and bosonic degrees of freedom in the original supersymmetric theory and is known as
dimensional regularization by dimensional  reduction \cite{Siegel:1979wq}\footnote{Note that for the $D=4$ and $\mathcal{N}=1$
or $\mathcal{N}=2$ super Yang-Mills theories this would amount to include $S=2\ve$ and $S=2+2\ve$ scalars respectively.}.

For the computation of correlation functions of the scalar fields $\phi^{a}_{i}(x)$ to next-to-leading order, i.e.~$\cO(g^2)$, we need to consider the inverse map $T^{-1}_{g}$
 of the interacting scalar fields $\phi^{a}_{i}(x)$ to the free-field correlators
\be\label{Scalar2point}
\big\langle \!\!\big\langle \phi_{i_{1}}^{a_{1}}(x_{1}) \phi_{i_{2}}^{a_{2}}(x_{2})\big\rangle\!\!\big\rangle \,=\,
\big\langle T_g^{-1}[\phi_{i_{1}}^{a_{1}}](x_{1})T_g^{-1}[\phi_{i_{2}}^{a_{2}}](x_{2}) \big\rangle_0
\ee
as well as 
\bea\label{Scalar4point}
\big\langle \!\!\big\langle \phi_{i_{1}}^{a_{1}}(x_{1}) \phi_{i_{2}}^{a_{2}}(x_{2})
\phi_{i_{3}}^{a_{3}}(x_{3})\phi_{i_{4}}^{a_{4}}(x_{4})\big\rangle\!\!\big\rangle \,&=&\,    \\[1mm]
&& \hspace{-4cm} =\,
 \big\langle T_g^{-1}[\phi_{i_{1}}^{a_{1}}](x_{1})T_g^{-1}[\phi_{i_{2}}^{a_{2}}](x_{2}) 
T_g^{-1}[\phi_{i_{3}}^{a_{3}}](x_{3})T_g^{-1}[\phi_{i_{4}}^{a_{4}}](x_{4}) \big\rangle_0   \nn
\, .
\eea

For $\cN=4$ super Yang-Mills theory, the action of inverse map $T^{-1}_{g}$ on the vector 
fields $A_\mu^a$ and the six scalar fields $\phi_i^a$  is easily derived by applying the split
(\ref{dimred})  to the formula (\ref{Tginv}). For the gauge field this gives
\begin{align}\label{Tginvgauge}
(T_g^{-1} A)^a_\mu(x) \,&=\,  A_\mu^a(x) \, - \, g f^{abc} \int du\, \pa_\l C(x-u) A_\mu^b(u) A_\l^c(u)  \nl
& \; + \,\frac12 g^2 f^{abc} f^{bde} \int dv dw\, 
    \Big[  -\pa_\r C(x-v) A_\s^c(v) \pa_\s C(v-w) A_\r^d(w) A_\mu^e(w) \nl
&  \qquad\qquad+ \pa_\r C(x-v) A_\s^c(v) \pa_\r C(v-w) A_\s^d(w) A_\mu^e(w) \nl
&  \qquad\qquad+ \pa_\r C(x-v) \phi_j^c(v) \pa_\r C(v-w) \phi_j^d(w) A_\mu^e(w) \nl
& \qquad\qquad   -\pa_\r C(x-v) A_\s^c(v) \pa_\mu C(v-w) A_\s^d(w) A_\r^e(w) \nl
& \qquad\qquad   -\pa_\r C(x-v) \phi_j^c(v) \pa_\mu C(v-w) \phi_j^d(w) A_\r^e(w) \nl
& \qquad\qquad + 2 \,\pa_\r C(x-v) A_\mu^c(v) \pa_\s C(v-w) A_\s^d(w) A_\r^e(w) \nl
& \qquad\qquad  -2 \, \pa_\r C(x-v) A_\r^c(v) \pa_\s C(v-w) A_\s^d(w) A_\mu^e(w) \Big]
  \; + \;\cO(g^3)
\end{align}
while for the scalar fields we obtain
\begin{align}
\label{Tinvscalar}
(T_g^{-1}\phi)^a_i(x) \,=\,  \phi_i^a(x) \, - \, g f^{abc}  &\int d^{D}u\, \pa_\l C(x-u) \phi_i^b(u) A_\l^c(u)
 \\[2mm]
   + \, \frac{g^2}{2} f^{abc}f^{bde} \int d^{D}u d^{D}v \,
 \Bigl [ 
 & -  \pa_\r C(x-u) A_\l^c(u) \pa_\l C(u-v) A_\r^d (v) \phi_i^e(v)  \nonumber \\ 
 & +  \pa_\r C(x-u) A_\l^c(u) \pa_\r C(u-v) A_\l^d (v) \phi_i^e(v)  \nonumber \\ 
 & +  \pa_\r C(x-u) \phi_j^c(u) \pa_\r C(u-v) \phi_j^d (v) \phi_i^e(v)  \nonumber \\ 
 & + 2  \pa_\r C(x-u) \phi_i^c(u) \pa_\l C(u-v) A_\l^d (v) A_\r^e(v)  \nonumber \\ 
 & - 2  \pa_\r C(x-u) A_\r^c(u) \pa_\l C(u-v) A_\l^d (v) \phi_i^e(v) \, \Bigr ] 
\, + \, \cO(g^3)\, . \nonumber
\end{align}
In the calculation we will apply regularization by dimensional reduction with 
\be
D=\delta^{\mu}_{\mu}=4-2\epsilon \;\;,\quad  S=\delta^{i}_{i}=6+2\epsilon
\ee
This implies $D+S=10$ which in fact is the combination
always arising in our computations up to one-loop order. 
The $D$-dimensional scalar propagator in position space reads
\be
C(x):= \int \frac{d^{2\omega}p}{(2\pi)^{2\omega}} \frac{1}{p^{2}} e^{ip\cdot x}
=\frac{\Gamma(\omega-1)}{4\pi^{\omega}}\, [x^{2}]^{1-\omega}\, .
\ee
In particular, we set the self-contraction $C(0)=0$ as a consequence of dimensional regularization for scale-less
integrals.

\subsection{Two-point function}

To compute the scalar two-point function to $\cO(g^2)$ we insert the expansion (\ref{Tinvscalar}), retaining
only terms of order $\cO(g^2)$ (the first order contributions vanish trivially), and then perform
the necessary Wick contractions. Just as before in the computation of the gauge fields in (\ref{2point}) there
are two contributions to this correlator: The contractions of the $\cO(g)$ terms emerging from
each operator $T_g^{-1}[\phi_{i}^{a}]$
 as well as the contractions of the $\cO(g^{2})$
terms of one $T^{-1}_{g}[\phi_{i}^{a}]$ with the leading $\phi_{i}^{a}$ term of the other.
 A straightforward calculation gives the result 
\begin{align}
\big\langle T_g^{-1}[\phi_{i_{1}}^{a_{1}}](x)T_g^{-1}[\phi_{i_{2}}^{a_{2}}](y) \big\rangle_0 &= \delta_{i_{1}i_{2}}\, \delta^{a_{1}a_{2}}\, C(x-y) \, \nonumber \\
 + g^{2} N \, \delta_{i_{1}i_{2}}\, \delta^{a_{1}a_{2}}\int d^{D}u\, d^{D}v \Bigl \{ &
\left (\frac{D+S}{2}-2\right )\, [ \, C(x-u) C(u-v) \pa_{\r} C(y-v)\pa_{\r}C(v-u)
+ (x\leftrightarrow y)\, ] \nonumber\\
& + C(u-v)^{2}\, \pa_{\r}C(x-u) \, \pa_{\r}C(y-v)\,
\Bigr \} +\cO(g^{4})\, .
\label{48eq}
\end{align}
where the $\cO(1) \times \cO(g^{2})$ contractions yield the term proportional to $(\frac12(D+S)-2)$ while the
$\cO(g)\times \cO(g)$ contractions yield the second term in the above.
Importantly, in the course of performing the $\cO(1)\times \cO(g^{2})$ Wick-contractions
one also takes into account the self-contractions of the $\cO(g^{2})$ terms, i.e.~the
operator insertions $T_g^{-1}[\phi_{i}^{a}]$ are not to be understood as normal ordered.
 
All integrals appearing in (\ref{48eq}) may in fact be reduced to the bubble integral 
\be
\raisebox{-0.5cm}{\includegraphics{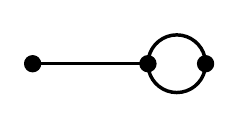}}=\int d^{D} u \, C(x-u)\, C(y-u)^{2}=
 I_{xy}
\ee
which is  symmetric in $(x\leftrightarrow y)$ and which appeared already in (\ref{2pointA}). 
To see the symmetry, we integrate by parts, using $\Box C(x)=-\delta (x)$ to obtain the
integral relations
\begin{align}
\label{intrel2pt}
\int  d^{D}u\, d^{D}v \,  C(x-u) C(u-v) \pa_{\r} C(y-v)\pa_{\r}C(v-u) & = -\frac{1}{2}\, I_{xy}
\nn \\
\int  d^{D}u\, d^{D}v\, 
C(u-v)^{2}\, \pa_{\r}C(x-u) \, \pa_{\r}C(y-v)& = I_{xy} \, .
\end{align}
The complete result for the two-point scalar correlation function in $\mathcal{N}=4$ super Yang-Mills
theory up to one-loop accuracy therefore reads
\be
\big\langle \!\!\big\langle \phi_{i_{1}}^{a_{1}}(x_{1}) \phi_{i_{2}}^{a_{2}}(x_{2})\big\rangle\!\!\big\rangle \,=\, \delta_{i_{1}i_{2}}\, \delta^{a_{1}a_{2}}\, 
C(x-y)\left(1 -2\, g^{2} N \, \frac{I_{xy}}{C(x-y)} \, \right )\, +\cO(g^{4}) \, ,
\ee
reproducing established results in the literature, see e.g.~\cite{Kristjansen:2002bb,Drukker:2008pi}.
We note that 
\be
\frac{I_{xy}}{C(x-y)}= \frac{1}{16\pi^{\omega}(2-\omega)}\, [(x-y)^{2}]^{2-\omega}
\quad \text{with $D=2\omega$}\, ,
\ee
yielding the expected logarithmic divergence near $D=4$.

\subsection{Four-point functions}
\label{4ptsec}

For the computation of the four-point function 
\be
\label{413}
\Big\langle T_g^{-1}[\phi_{i_{1}}^{a_{1}}](x_{1})T_g^{-1}[\phi_{i_{2}}^{a_{2}}](x_{2})  T_g^{-1}[\phi_{i_{3}}^{a_{3}}](x_{3})T_g^{-1}[\phi_{i_{4}}^{a_{4}}](x_{4}) \Big\rangle_0
\ee
we proceed from (\ref{Tinvscalar}). When expanding out
this formula to $\cO(g^{2})$ the combinatorics of Wick contractions grows considerably. 
Again one inserts $\cO(g)$ terms twice or $\cO(g^{2})$ terms once next to the leading
terms in the other slots. Here there are two types of color structures emerging: the connected 
terms are proportional to two structure constants $f^{a_{i}a_{j}e}f^{a_{k}a_{l}e}$ while the
disconnected terms are proportional to $\delta^{a_{i}a_{j}}\delta^{a_{k}a_{l}}$.
 
Gathering the connected terms one encounters the following integral identities
\begin{align}
\int du\, dv\, C(x_{1}-u)\, C(x_2-u)\, C(u-v)\, \partial_{\mu}C(x_{3}-v)\, 
\partial^{\mu}C(x_{4}-v)
&= \partial_{3}\cdot\partial_{4}H_{12;34} \\
\int du\, dv\,C(x_{1}-u)\, C(x_{3}-v)\, C(x_{4}-v)\, \partial_{\mu}C(x_{2}-u)\, \partial^{\mu}C(u-v)
&= - C_{12}Y_{234} + \partial_{1}\cdot\partial_{2}H_{12;34} \nn
\end{align}
where we defined the $H$ and $Y$-functions
\begin{align}
\label{intsHY}
\raisebox{-0.5cm}{\includegraphics[scale=.6]{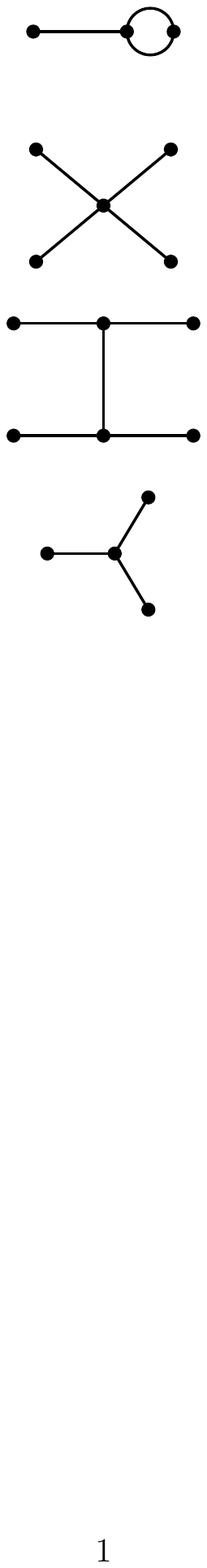}}&=\int d u\, dv\,  
 C(x_{1}-u) 
 C(x_{2}-u)  C(u-v) C(x_{3}-v)  C(x_{4}-v) \,\equiv \, H_{12,34}\, , \nn\\
 \raisebox{-0.5cm}{\includegraphics[scale=.7]{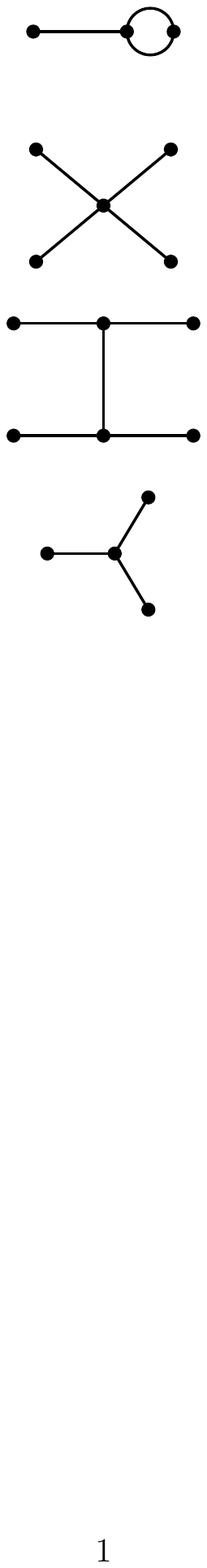}}&=\int d u\,  
 C(x_{1}-u) 
 C(x_{2}-u)  C(x_{3}-u)  \,\equiv \, Y_{123}\, . 
\end{align}
The connected part may then be brought into the form (with $C_{12} \equiv C(x_1-x_2)$)
\begin{align}
&(\ref{413})_{\text{connected}}= f^{a_{1}a_{2}e}f^{a_{3}a_{4}e}\, \Bigl [ \delta_{i_{1}i_{2}}
 \delta_{i_{3}i_{4}}\, (\partial_{1}-\partial_{2})\cdot(\partial_{3}-\partial_{4})\, H_{12;34}
 \nn\\
 &+ (\delta_{i_{1}i_{3}}\delta_{i_{2}i_{4}} -\delta_{i_{1}i_{4}} \delta_{i_{2}i_{3}}) 
 \Bigl \{ (\partial_{1}\cdot\partial_{2}+\partial_{3}\cdot
 \partial_{4})H_{12;34}- \frac{1}{2} C_{12}(Y_{134}+Y_{234})
 - \frac{1}{2} C_{34}(Y_{123}+Y_{124})
 \Bigr \} \Bigr ]\nn\\
 &+ \text{permutations}
 \end{align}
The integrals in the last line may be reduced upon noting that 
\be
\partial_{1}\cdot\partial_{2}H_{12;34} \,=\,  \frac{1}{2} C_{12} ( Y_{134}+Y_{234})\, - \,
\frac{1}{2} X_{1234}
\ee
using partial integrations and where we introduced the $X$ integral
\be
\raisebox{-0.5cm}{\includegraphics[scale=.6]{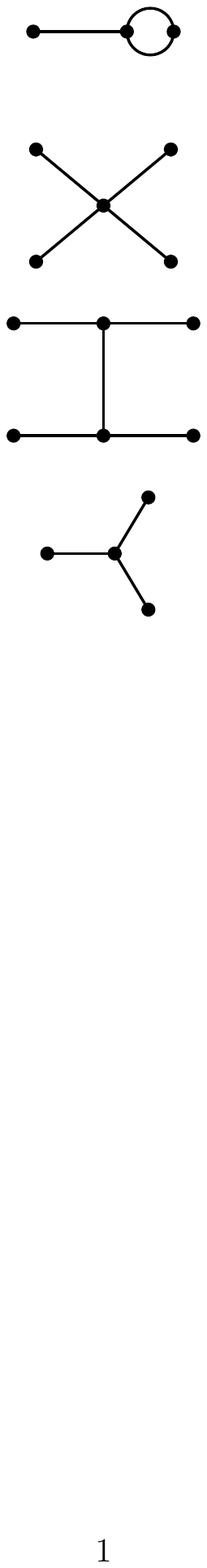}}=\int d u\, C(x_{1}-u) 
 C(x_{2}-u)  C(x_{3}-u)  C(x_{4}-u) = X_{1234} \, .
\ee
The $H,Y$ and $X$ integrals are also known analytically, {\em cf.}~\cite{Beisert:2002bb}. 
Putting everything together we obtain the connected part of four-point function
up to $\cO(g^{2})$ 
\begin{align}
\label{4ptcon}
\big\langle \!\!\big\langle \phi_{i_{1}}^{a_{1}}(x_{1}) & \phi_{i_{2}}^{a_{2}}(x_{2})
\phi_{i_{3}}^{a_{3}}(x_{3})\phi_{i_{4}}^{a_{4}}(x_{4})\big\rangle\!\!\big\rangle_{\text{connected}} = \\
&\phantom{+} g^{2}f^{a_{1}a_{2}e}\, f^{a_{3}a_{4}e}\, \left [ \delta_{i_{1}i_{2}}\delta_{i_{3}i_{4}}
 (\pa_{1}-\pa_{2})\cdot  (\pa_{3}-\pa_{4})\,H_{12,34}   {-} 
 (\delta_{i_{1}i_{3}}\delta_{i_{2}i_{4}} - \delta_{i_{1}i_{4}}\delta_{i_{2}i_{3}})
 \, X_{1234} \right ] \nn\\
 &+g^{2}f^{a_{1}a_{3}e}\, f^{a_{2}a_{4}e}\, \left [ \delta_{i_{1}i_{3}}\delta_{i_{2}i_{4}}
 (\pa_{1}-\pa_{3})\cdot  (\pa_{2}-\pa_{4})\,H_{13,24}   {-} 
 (\delta_{i_{1}i_{2}}\delta_{i_{3}i_{4}} - \delta_{i_{1}i_{4}}\delta_{i_{2}i_{3}})
 \, X_{1234} \right ] \nn\\
 &+g^{2}f^{a_{1}a_{4}e}\, f^{a_{2}a_{3}e}\, \left [ \delta_{i_{1}i_{4}}\delta_{i_{2}i_{3}}
 (\pa_{1}-\pa_{4})\cdot  (\pa_{2}-\pa_{3})\,H_{14,23}   {-} 
 (\delta_{i_{1}i_{2}}\delta_{i_{3}i_{4}} - \delta_{i_{1}i_{3}}\delta_{i_{2}i_{4}})
 \, X_{1234} \right ] \nn\\
 & + \cO(g^{4})\, . \nn
\end{align}
For the disconnected terms one consistently finds pairwise appearances of two-point function
contractions
\begin{align}
\label{4ptdis}
\big\langle \!\!\big\langle & \phi_{i_{1}}^{a_{1}}(x_{1})  \phi_{i_{2}}^{a_{2}}(x_{2})
\phi_{i_{3}}^{a_{3}}(x_{3})\phi_{i_{4}}^{a_{4}}(x_{4})\big\rangle\!\!\big\rangle_{\text{disconnected}} = \\[1mm] &
\delta^{a_{1}a_{2}}\delta^{a_{3}a_{4}}\delta_{i_{1}i_{2}}\delta_{i_{3}i_{4}} 
\big(C(x_{1}-x_{2}) -2\, g^{2}N I_{12}\big) \big(C(x_{3}-x_{4})
-2 \, g^{2} N I_{34}\big) + \text{permutations}+ \cO(g^{4})\, , \nn
\end{align}
using the identical integral relations of (\ref{intrel2pt}).
These results of (\ref{4ptcon}) and (\ref{4ptdis}) reproduce the known results 
obtained in the standard perturbative computation in $\mathcal{N}=4$ super Yang-Mills theory, see {\em e.g.}~\cite{Drukker:2008pi}. In consequence, \emph{any} $n$-point scalar field correlation
function will be reproduced up to the $\cO(g^{2})$ order using the inverse non-local map
$T_{g}^{-1}$ to the free gauge theory. This is due to the fact that the connected part of $n$-point scalar correlators are of order $\cO(g^{n-2})$. Hence, at order $\cO(g^{2})$ accuracy only
the disconnected parts will contribute for $n>4$. 

\section{Deriving the one-loop dilatation operator}

A central class of gauge invariant observables in $\mathcal{N}=4$ super Yang-Mills theory
are the anomalous scaling dimensions of composite operators. They have been 
subject to intense studies and remarkable
results were produced 
in the AdS/CFT integrability program, including exact results to all orders in 
$g^{2}N$  in the planar $N\to\infty$
limit of the $SU(N)$ gauge theory \cite{Beisert}. Focusing 
on the class of composite operators built from scalar fields, these 
are constructed as traces of the scalar fields $\phi_{i}(x)\equiv t^{a}\phi^{a}_{i}(x)$
at a common space-time point (with the $SU(N)$  generators $t^a$). 
These take the schematic form
\be
\cO_{\alpha} = \text{Tr}\,\big(\phi_{i} \phi_{j}\phi_{k}\ldots \big) \, 
\text{Tr}\,\big( \phi_{l}\phi_m \phi_n \ldots \big) \, \cdots
\ee
where $\alpha$ is a superindex labeling all possible compositions.
As a consequence of the conformal symmetry the two-point functions of these (renormalized)  operators take canonical form
\be
\big\langle\!\!\big\langle \cO^{\text{ren}}_{\alpha}(x)\, \cO^{\text{ren}}_{\beta}(0) 
\big\rangle\!\!\big\rangle
= \frac{\delta_{\alpha\beta}}{[x^{2}]^{\Delta_{\alpha}(g^{2},N)}
}\, , 
\ee
where the scaling dimensions receive perturbative corrections in an expansion in $g^{2}$
starting out with the  the naive classical (tree level) scaling dimension 
obtained by standard power counting. In order to achieve this the operator mixing
problem needs to be resolved.
A superior tool for doing this (and thereby finding the $\Delta_{n}(g^{2},N)$)
is the construction of the dilatation operator $\hat D$ as developed in \cite{Beisert:2003tq},
following initial results at one-loop in \cite{Minahan:2002ve,Beisert:2002ff}. 
The dilatation operator $\hat D$ acts on
states at the origin of space-time (in a radial quantization scheme) -- its eigenvalues
correspond to the anomalous dimensions 
\be
\hat D\,  \cO_{\alpha} = \Delta_{\alpha}(g^{2},N)\, \cO_{\alpha} \,.
\ee
We now wish to extract the dilatation operator from our inverse map $T_{g}^{-1}$ to
order $g^{2}$ by taking the two-point limit of the four-point results in section \ref{4ptsec}.
For this we need to establish some technology following \cite{Beisert:2003tq}.
To begin with, we distinguish the fields at points $x$ and $0$ by the superscript $\pm$
\be
\Phi^{+}_{i}\,\equiv\, t^{a}\, \phi^{a}_{i}(x)\, , \quad
\Phi^{-}_{i} \, \equiv \, t^{a}\, \phi^{a}_{i}(0)\, . 
\ee
The tree-level two point function of composite 
operators at these points may then formally be written as
\be
\label{W0define}
\langle\!\langle \cO_{\alpha}^{+}\,\cO_{\beta}^{-} \rangle\!\rangle_{\text{tree}} =
\exp [ W_{0}(x, \cPhi^{+},\cPhi^{-})]\, \cO^{+}_{\alpha}\cO^{+}_{\beta}\Bigr |_{\Phi=0}
\ee
with the field derivatives
\be
\cPhi_{i}^{+}:= t^{a} \frac{\delta}{\delta \phi^{a}_{i}(x)}\, , \quad
\cPhi_{i}^{-}:= t^{a} \frac{\delta}{\delta \phi^{a}_{i}(0)}\, , \quad
\ee
and the tree level generator $W_{0}$ inserting free field scalar propagators in between
$\Phi^{+}_{i}$ and $\Phi^{-}_{i}$ 
\be
W_{0}(x, \cPhi^{+},\cPhi^{-})= C(x)\, \Tr \, \cPhi^{+}_{i}\cPhi^{-}_{i}\, .
\ee
The exponentiated $W_{0}$ in equation \ref{W0define} performs free Wick contractions between
all constituent fields of the $\cO^{+}_{\alpha}$ and $\cO^{-}_{\beta}$ and thus computes
the tree-level correlator. The one-loop
correction to this two-point correlator then takes the form
\be
\label{W1define}
\langle \!\langle \cO_{\alpha}^{+}\,\cO_{\beta}^{-} \rangle\!\rangle_{\text{one-loop}} =
\exp [ W_{0}(x, \cPhi^{+},\cPhi^{-})]\,
\left(1+ g^{2}\, W_{2}(x, \cPhi^{+},\cPhi^{-})\right)
 \cO^{+}_{\alpha}\cO^{+}_{\beta}\Bigr |_{\Phi=0} \, .
\ee
Let us now extract  $W_{2}(x, \cPhi^{+},\cPhi^{-})$ from the
pinching limits of our four-point functions  \eqn{4ptcon} and \eqn{4ptdis} by taking $x_{1,2}\to x$ and
$x_{3,4}\to 0$. In order to do this we need the
 following integral identities obtained by standard one-loop Feynman integral
techniques in dimensional regularization
\begin{align}
X_{00xx}= \raisebox{-0.24cm}{\includegraphics[scale=.6]{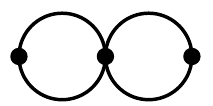}}&=\int du\,  C(x-u)^{2}C(u)^{2}= 2 \, \Omega(x) \, C(x)^{2} +\cO(\epsilon) \nn \\
Y_{00x}=Y_{0xx}=I_{x0}&= \int du\, C(x-u)^{2}C(u)=\Omega(x) \, C(x)^{2} +\cO(\epsilon)
\end{align}
with the common divergent factor 
\be
\Omega(x)= \frac{1}{16\pi^{\omega}}\frac{1}{2-\omega}
[x^{2}]^{2-\omega} \, , \qquad D=2\omega=4-2\epsilon
\ee
 and the scalar propagator $C(x)=\frac{\Gamma(\omega-1)}{4\pi^{\omega}}
[x^{2}]^{1-\omega}$. More subtle are the pinching limits of the derivatives of the
$H$ functions appearing in \eqn{intsHY} which amount to two-loop Feynman integrals. 
Defining the relevant integral as
\be
\tilde H_{12;34}= (\partial_{1}-\partial_{2})\cdot (\partial_{3}-\partial_{4}) H_{12;34}
\ee
the key identities we found are
\begin{align}
\tilde H_{00;xx}&= 0 \nn\\
\tilde H_{0x;0x}&= 2 \, \Omega(x) \, C(x)^{2} +\cO(\epsilon) \, .
\end{align}
Whereas the first relation is easy to see, reaching the second relation we made use of the
{\tt Tarcer} package \cite{Mertig:1998vk}.  We also cross checked our result with the results of \cite{Beisert:2003tq} in Appendix B. Concretely the identities established are
\begin{align}
\raisebox{-0.5cm}{\includegraphics[scale=.6]{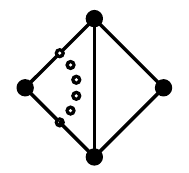}}&=\int du\, dv \, \partial_{\mu} C(x-u)\, \partial_{\mu} C(x-v) \, C(u-v) \, C(y-u)\, C(y-v)\nn\\
&= \Omega(x-y) \, C(x-y)^{2} +\cO(\epsilon) \nn \\
\raisebox{-0.5cm}{\includegraphics[scale=.6]{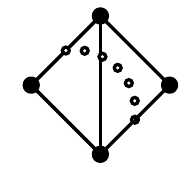}}&=\int du\, dv \, \partial_{\mu} C(x-u)\,  C(x-v) \, C(u-v) \, 
C(y-u) \partial_{\mu}C(y-v)\nn\\ &= (2-\omega)\, \Omega(x-y)\, C(x-y)^{2} + \cO(\epsilon^{2}) \, ,
\end{align}
where the dots on the graphs indicate the contracted indices of the derivatives.
Importantly the second integral is not divergent and does not contribute to the 
anomalous dimensions.
Using theses results the connected part of the pinched four point function of \eqn{4ptcon} in the limit $x_{1,2}\to x$ and $x_{3,4}\to 0$ takes the form
\begin{align}
\label{2ptcon}
\big\langle \!\!\big\langle \phi_{i_{1}}^{a_{1}}(x)  \phi_{i_{2}}^{a_{2}}(x)
\phi_{i_{3}}^{a_{3}}(0)&\phi_{i_{4}}^{a_{4}}(0)\big\rangle\!\!\big\rangle_{\text{connected}} = 
\nn\\ 
 = \, 2\, \Omega(x)\, C(x)^{2}\, g^{2}  
&\Bigl ( f^{a_{1}a_{2}e}\, f^{a_{3}a_{4}e}\, \left [ \delta_{i_{1}i_{4}}\delta_{i_{2}i_{3}}-
 \delta_{i_{1}i_{3}}\delta_{i_{2}i_{4}} 
  \right ] \nn\\
 &+f^{a_{1}a_{3}e}\, f^{a_{2}a_{4}e}\, \left [ \delta_{i_{1}i_{3}}\delta_{i_{2}i_{4}}
  - 
 \delta_{i_{1}i_{2}}\delta_{i_{3}i_{4}} + \delta_{i_{1}i_{4}}\delta_{i_{2}i_{3}}
  \right ] \nn\\
 &+f^{a_{1}a_{4}e}\, f^{a_{2}a_{3}e}\, \left [ \delta_{i_{1}i_{4}}\delta_{i_{2}i_{3}}
-
 \delta_{i_{1}i_{2}}\delta_{i_{3}i_{4}} + \delta_{i_{1}i_{3}}\delta_{i_{2}i_{4}}
  \right ] \Bigr ) \, .
\end{align}
Using the matrix variation notation introduced above this correlator may be 
translated to the operator\footnote{Our conventions are $[t^{a},t^{b}]=if^{abc}t^{c}$ and 
$\Tr \, t^{a}t^{b}=\delta^{ab}$, we also note $f^{aef}f^{bef}=N \delta^{ab}$.}
\begin{align}
\label{W2C}
W_{2}^{C}=  \Omega(x)\, C(x)^{2}\Bigl \{ &
\Tr [\cPhi^{+}_{i},\cPhi^{+}_{j}]\, [\cPhi^{-}_{i},\cPhi^{-}_{j}]
+\frac{1}{2}\, \Tr [\cPhi^{+}_{i},\cPhi^{-}_{j}]\, [\cPhi^{+}_{i},\cPhi^{-}_{j}] \nn \\
& - 2\, \Tr [\cPhi^{+}_{i},\cPhi^{+}_{i}]\, [\cPhi^{-}_{j},\cPhi^{-}_{j}] \, \Bigr \}
\end{align}
which acts on the  two operators $\cO^{+}_{\alpha}\cO^{-}_{\beta}$ located at $x$ and $0$.
It yields the correlation function in the sense of \eqn{W1define}. The disconnected
contribution arises from \eqn{4ptdis} and takes the form 
\begin{align}
\big\langle \!\!\big\langle \phi_{i_{1}}^{a_{1}}(x)  \phi_{i_{2}}^{a_{2}}(x)
\phi_{i_{3}}^{a_{3}}(0)&\phi_{i_{4}}^{a_{4}}(0)\big\rangle\!\!\big\rangle_{\text{disconnected}} =\nn\\
&  (\delta_{i_{1}i_{3}}\delta_{i_{2}i_{4}}\delta^{a_{1}a_{3}}\delta^{a_{2}a_{4}}
+ \delta_{i_{1}i_{4}}\delta_{i_{2}i_{3}}\delta^{a_{1}a_{4}}\delta^{a_{2}a_{3}})
\Bigl \{ 1-4\, g^{2}N\, \Omega(x)\Bigr \} \, ,.
\end{align}
Translating the one-loop $\cO(g^{2})$ contribution to the matrix variation notation yields 
\be
\label{W2D}
W_{2}^{D}= 2 \, \Omega(x)\, C(x)^{2}
\, \Tr[t^{a},\cPhi^{+}_{i}]\, [t^{a},\cPhi^{-}_{i}]\, .
\ee
In order to extract the dilatation operator from these results we return to
\eqn{W1define} and now change the argument of $W_{2}$ from
$\cPhi^{+}$  to $C(x)^{-1} \Phi^{-}$ at the cost of normal ordering the  $W_{2}$ 
\be
\label{W1define2}
\langle \cO_{\alpha}^{+}\,\cO_{\beta}^{-} \rangle_{\text{one-loop}} =
\exp [ W_{0}(x, \cPhi^{+},\cPhi^{-})]\,
\left(1+ g^{2}\, V_{2}(x)\, \right)
 \cO^{+}_{\alpha}\cO^{+}_{\beta}\Bigr |_{\Phi=0} \, .
\ee
with the normal ordered one loop effective vertex \cite{Beisert:2003tq}
\be
V_{2}(x) = :W_{2}(x,C(x)^{-1} \Phi^{-},\Phi^{-}): \, .
\ee
This replacement may be done, as the result $\vev{\cO_{\alpha}\cO_{\beta}}$
vanishes unless every $\Phi^{-}$ is contracted with a $\Phi^{+}$ before the fields are
set to zero. Here, the only possibility is to contract with a term in $W_{0}$ which
effectively changes the argument back to $\cPhi^{+}$. Normal ordering $:\,:$
secures that no
new contractions are introduced within $W_{2}$. Operator renormalization is then performed via
\be
\cO^{\text{ren}}= \left (1-\sfrac{1}{2}g^{2}V_{2}(x_{0})\right ) \cO
\ee
with an arbitrary reference point $x_{0}$. The resulting two-point function is finite
\be
\label{W1ren}
\langle \cO_{\alpha}^{\text{ren}\,+}\,\cO_{\beta}^{\text{ren}\, -} \rangle_{\text{one-loop}} =
\exp [ W_{0}(x, \cPhi^{+},\cPhi^{-})]\,
\left(1+ g^{2}\, V_{2}(x)- g^{2}\, V_{2}(x_{0})\, \right)
 \cO^{+}_{\alpha}\cO^{+}_{\beta}\Bigr |_{\Phi=0} \, .
\ee
The dilatation operator $D_{2}$ is now extracted upon sending the regulator to zero
\be
\lim_{\epsilon\to 0} \, \big(V_{2}(x)-V_{2}(x_{0})\big) = \log(x_{0}^{2}/x^{2})\, \hat D_{2}
\ee
with 
\be
\hat D_{2}=- \lim_{\epsilon\to 0} \epsilon\, V_{2}(x)
\ee
as the $\log x^{2}$ contribution to $V_{2}(x)$ is always paired with the  $1/\epsilon$ pole in dimensional regularization. The final answer for the renormalized two-point
function then reads
\be
\label{W1ren2}
\langle \cO_{\alpha}^{\text{ren}\,+}\,\cO_{\beta}^{\text{ren}\, -} \rangle_{\text{one-loop}} =
\exp [ W_{0}(x, \cPhi^{+},\cPhi^{-})]\,
\exp[g^{2}\log(x_{0}^{2}/x^{2})\, D_{2}]
 \cO^{+}_{\alpha}\cO^{+}_{\beta}\Bigr |_{\Phi=0} \, .
\ee
Applying this rational to our results \eqn{W2C} and \eqn{W2D} we find
\be
\hat D_{2}=-\frac{1}{8\pi^{2}}\, \left ( 
:\Tr [\Phi_{i},\Phi_{j}]\, [\cPhi_{i},\cPhi_{j}]:
-\sfrac{1}{2}\, :\Tr [\Phi_{i},\cPhi_{j}]\, [\Phi_{i},\cPhi_{j}]: \right )
+  \sfrac{1}{8\pi^{2}}\,\hat V_{D}
\ee
where we separated off the piece $V_{D}$ which turns out to just amount to a gauge
transformation generated by $\hat G^{a}= \Tr [t^{a},\Phi_{i}]\, \cPhi_{i}$ as
\begin{align}
\hat V_{D}&= :\Tr [\Phi_{i},\cPhi_{i}]\, [\Phi_{j},\cPhi_{j}]:
+ \, \Tr[\Phi_{i}, t^{a}]\, [t^{a},\cPhi_{i}] \nn \\
&=  \Tr [\Phi_{i},\cPhi_{i}]\, [\Phi_{j},\cPhi_{j}]= 
\Tr \left ([t^{a},\Phi_{i}] \cPhi_{i} \right )\, \Tr\left ( [ t^{a},\Phi_{j}]\cPhi_{j}\right )
 = \hat G^{a}\, \hat G^{a}\, . 
\end{align}
Hence $V_{D}$ vanishes on gauge invariant composite operators and the one-loop
dilatation operator in the scalar sector reads
\be
\hat D_{2}=-\sfrac{1}{8\pi^{2}}\, \left ( 
:\Tr [\Phi_{i},\Phi_{j}]\, [\cPhi_{i},\cPhi_{j}]:
-\sfrac{1}{2}\, :\Tr [\Phi_{i},\cPhi_{j}]\, [\Phi_{i},\cPhi_{j}]: \right ) \, .
\ee
It precisely coincides with the dilatation operator established in \cite{Beisert:2003tq} upon adapting the conventions for the gauge coupling constants. As a consequence \emph{all}
scaling dimensions in the scalar $SO(6)$ sector of $\mathcal{N}=4$ SYM are reproduced with the map $T_{g}$  up to the $g^{2}$ order.

\section{Outlook}
In principle there are now many topics to explore in terms of the map $T_g$. 
One especially interesting question is how the present formalism 
applies to the computation of the Wilson loop integral
\be\label{WL1}
\big\langle \!\!\big\langle  W \left( \cC\right) \big\rangle\!\!\big\rangle  \,\equiv \,
\bigg\langle \!\!\!\!\bigg\langle  \cP \exp\left( ig \oint_\cC A^a_\mu t^a  dx^\mu \right) \bigg\rangle\!\!\!\!\bigg\rangle
\ee
with $\cC$ a closed curve in $\mathbb{R}^4$ and fundamental $SU(N)$ generators 
$t^a$. In principle we can evaluate this by considering
\be\label{WL2}
\left\langle  \cP \exp\left( ig \oint_\cC (T_g^{-1}A) ^a_\mu t^a  dx^\mu \right) \right\rangle_0
\ee
which again can be determined up to $\cO(g^2)$ for special cases of interest,  making
use of the results of the previous chapters. As the $n$-point correlators agree to
$\cO(g^{2})$, as was shown, the perturbative evaluation of the Wilson loop (\ref{WL2}) using the inverse 
map $T^{-1}_{g}$  is guaranteed to reproduce the original expectation  value (\ref{WL1}) using
standard perturbation theory to that order. 

An interesting extension of the above lies in the study of supersymmetric Maldacena-Wilson
loops \cite{Maldacena:1998im}. Here the path couples to the gauge fields and the scalars,
i.e.~the loop exponent takes the form
\be
 W_{S}(\cC)= \cP \exp\left( ig \int_0^{1} (A^a_\mu   {\dot x}^\mu  + i\phi^{a}_{i} |\dot x |\, \theta^{i}) t^a ds \right) 
\quad \text{with} \quad \theta^{i}\theta^{i}=1\, ,
\ee
where we have parametrized the loop $\cC$ by $x^\mu = x^\mu(s)$ with $0\leq s \leq 1$.
For special curves such as a straight line or a circle the Maldacena-Wilson loop
expectation value $\langle\! \langle W_{S}(\cC) \rangle\!\rangle$ \emph{does not} receive contributions from bulk interactions \cite{Erickson:2000af,Drukker:2000rr} in a Feynman
diagrammatic evaluation. Put differently it is equal to the same Wilson loop operator
in the \emph{free} gauge theory
\be
\big\langle \!\!\big\langle  W_{S} \left( \cC\right)  \big\rangle\!\!\big\rangle 
= \big\langle  W_{S} \left( \cC\right) \big\rangle_{0} \, .
\ee
Hence from the perspective of this work for these special geometries the Maldacena-Wilson loop operator should 
be \emph{invariant} under the $\cR$ map.
Evaluating the $\cR$ on (\ref{WL1}) we deduce with  (\ref{cR}) (now interpreting the gauge fields
as 10 dimensional)
\begin{align}
\cR \, & \big\langle \!\!\big\langle  \, W \left( \cC\right) \big\rangle\!\!\big\rangle  = 
\big\langle \!\!\big\langle  W \left( \cC\right) \big\rangle\!\!\big\rangle \;  -
\nn\\[2mm]
&  \hspace{-5mm}  - \; \frac1{2r_D} \int_0^1 ds\,\int \, du\,dv\; 
\Pi_{\mu\nu}(x(s) -u){\rm Tr}\, \big( \c_\nu \c^{\r\s} S^{ba}(v-u)\big) f^{bcd} A_\r^c(v)A_\s^d(v)   \times \nl[1mm]
   \hspace{-1.5cm}  & \times
\Big\langle \!\!\!\Big\langle  \left( \cP \exp\left( ig \int_0^s dt  A_\mu(x(t))  \dot x^\mu(t) dt \right) \right)  
 t^a \dot x^\mu(s)  \left( \cP \exp\left( ig \int_s^1 dt  A_\mu(x(t))  \dot x^\mu(t) \right)\right)  
\Big\rangle\!\!\!\Big\rangle \,.
\end{align}
Note that the first term on the r.h.s. (resulting from the application of $d/dg$)
is again the Wilson loop operator. Invariance under $\cR$ thus 
amounts to the vanishing of the remaining expressions for special 
contours.
Working this out in detail is left to future work.

A central question concerns the existence of the $\cN=4$ theory {\em beyond} 
perturbation theory. As is well known, the construction of an interacting quantum 
field theory in {\em four} space-time dimensions obeying the Wightman axioms remains
an outstanding problem of quantum field theory  (see {\em e.g.} \cite{GJ}). In that framework, 
the non-triviality of the theory would be ensured by ascertaining the non-triviality of 
the $S$-matrix. Among all rigidly supersymmetric theories the $\cN=4$ theory would 
seem to come closest to realizing this quantum field theorist's dream. However,  being an exactly 
conformal theory without asymptotic one-particle states, it has no $S$-matrix in the usual sense.
Hence standard arguments do not apply; rather, it appears that the Wightman axioms of 
ordinary quantum field theory must be replaced by the axioms of the conformal bootstrap
program \cite{Rychkov}. A major goal of the present approach (and still a dream) would 
be to exploit the existence of the map $T_g$ and its properties towards a completely rigorous
construction of $\cN=4$ Yang-Mills theory at the non-perturbative level.

\vspace{0.4cm}

\noindent{\bf Acknowledgments:}  H.N. would like to thank S. Ananth, O. Lechtenfeld and
M.~Bochicchio for discussions and correspondence related to this work.

\end{document}